# Impact of Economic Constraints on the Projected Timeframe for Human-Crewed Deep Space Exploration

Philip E. Rosen[1], Dan Zhang[2], Jonathan H. Jiang[3], Leopold Van Ijzendoorn[4], Kristen A. Fahy[3], Zong-Hong Zhu[2]

[1.]Retired energy industry engineer, an independent researcher, WA 98662, USA
[2.]Department of Astronomy, Beijing Normal University, Beijing, China
[3.]Jet Propulsion Laboratory, California Institute of Technology, Pasadena, CA 91108, USA
[4.]Department of Computer Sciences, Radboud University, The Netherlands

Correspondence: Jonathan.H.Jiang@jpl.nasa.gov

## Abstract

Deep space exploration offers the most profound opportunity for the expansion of humanity and our understanding of the Universe, but remains extremely challenging. Progress will continue to be paced by uncrewed missions followed up by crewed missions to ever further destinations. Major space powers continue to invest in crewed deep space exploration as an important national strategy. An improved model based on previous work is developed, which projects the earliest possible launch dates for human-crewed missions from cis-lunar space to selected destinations in the Solar System and beyond based on NASA's historic budget trend and overall development trends of deep space exploration research. The purpose of the analysis is to provide a projected timeframe for crewed missions beyond Mars. Our findings suggest the first human missions from a spacefaring nation or international collaboration to the Asteroid Belt and Jovian System could be scheduled as soon as ~2071 to ~2087 and ~2101 to ~2121, respectively, while a launch to the Saturn System may occur by the year ~2132, with an uncertainty window of ~2129 to ~2153.

## 1. Introduction

On October 4, 1957, the former Soviet Union launched the world's first artificial satellite, Sputnik, into Earth orbit marking humanity's entry into the Space Age and the start of the "Space Race" with the United States. By 1961, human crewed exploration of near-space had commenced and a goal of landing on the Moon established. Spearheaded by the National Aeronautics and Space Administration (NASA), the United States was the first and thus far only nation to successfully accomplish human-crewed missions to the lunar surface, starting with Apollo 11 on July 20, 1969. The 1960s also saw the rapid advancement of deep space exploration by robotic probes as the first successful flyby missions to Venus, in 1962 by Mariner 2, and Mars, by Mariner 4 in 1964, were launched. Beyond cis-lunar space, however, only uncrewed missions covering the various celestial bodies in the solar system such as the other seven known planets, some of their satellites, the Sun, dwarf planets, asteroids, and comets [1-5] have been executed. The farthest of these space probes, Voyager 1, traversed the heliopause in 2012 as it exited the solar system for interstellar space. While most robotic probes' missions have entailed brief flyby type encounters with their objectives (e.g., outer planets by Pioneer 10 & 11 and Voyager 1 & 2, the dwarf planet Pluto and a Kuiper Belt object by New Horizons), others such as those of the Venera series (Venus), Viking 1 & 2 (Mars), Juno (Jupiter) and Cassini (Saturn) achieved orbit about their destinations, allowing for extended study. In addition, landers were included with some of these missions, returning images and detailed surface analysis to Earth-bound researchers.

Increasingly sophisticated robotic missions such as the Mars rovers will continue to play a critical role in deep space exploration by serving as data gathering instruments – but are themselves ultimately limited in autonomy and capability. If not as a prelude to human deep space exploration, and ultimately expansion, such activities may be considered as serving only scientific curiosity. With the Earth extensively explored and human population rushing past eight billion even as the resources of our otherwise bountiful home world continue to deplete, it is unsurprising that serious considerations are being given to colonizing off-world [6-8]. For an undertaking of this magnitude, and particularly as human lives would be put at risk in environments far more hostile than any found on Earth, careful planning is essential to success. Integral to detailed planning are the modeling of outcomes given the many constraints and influencing factors. A first step analysis towards predicting timeframes for first human-crewed launches from Earth to Solar System and interstellar destinations, while taking into account the anticipated pace of continued technological progress, was investigated by previous work [8]. However, as suggested in that study, how far and fast humanity reaches is a complex function of multiple variables and thus further analysis of imposed limits is needed for more precise conclusions to be drawn. This study will build upon the latter by accounting for the economic constraints associated with deep space exploration.

Deep space exploration activities not only promote humankind's research on the formation of the Solar System and the Universe, the origin and evolution of life as well as other important scientific issues, but also furthers the development of national economies, technology, and our civilization's drive towards a more enlightened society. Considering the unique challenges posed by direct human participation, missions which include crewed landings can serve to convincingly demonstrate scientific and technological strengths, as well as characterizing national competition. One need only recall the early 1960s when the USSR celebrated the first human flight into space and America's forceful response to send men to the Moon and return them safely to Earth before the close of that consequential decade. Major space powers continue to invest in crewed deep space exploration as an important national strategy [3-5], despite the high capital investment and operational costs. As a stand-out example, the United States invested 20.44 billion dollars (approximately 123.8 billion dollars, adjusted for inflation to year 2020 [15]) in the Apollo program alone. Deep space exploration, particularly where it would involve people undertaking long journeys, presents as a large-scale, complex scientific and technological project characterized by national/international scale investment, long development cycles, high human and technical risks, and challenges in organization and implementation. Since human spaceflight is still mainly funded by governments, although private investment is clearly on the rise, publicly derived and managed budgets will continue to influence the planned implementation of spaceflight programs.

As a working example, combining NASA's inflation-adjusted budget curve with the number of U.S. deep space missions launched (Figure 1) illustrates the logical cause and effect relationship between relatively larger budgets and more launches of such missions, after taking into consideration a common time lag between investment and return on research. In this paper an improved model of previous work [8] is developed based on NASA's budget trend and that of applicable technology development for human-crewed deep space exploration over the first six plus decades of the Space Age, from which is derived a projected timeframe of the earliest possible launch dates for such missions to selected destinations in the outer Solar System and is extendable to near interstellar space.

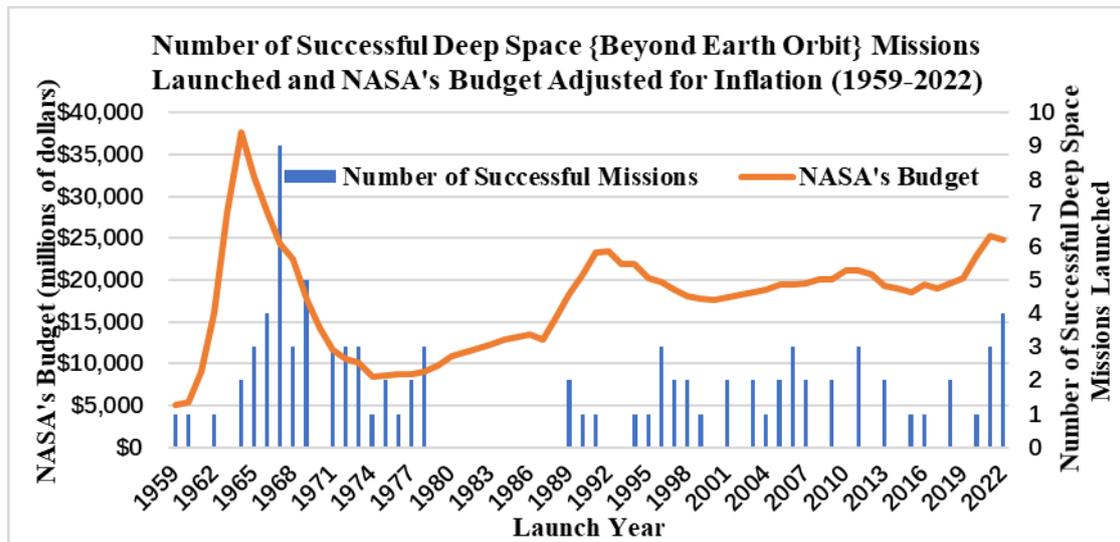

**Figure 1:** The orange line shows the change in NASA's budget from 1959 to 2021, adjusted for inflation. The blue columns show the number of United States deep space missions launched successfully from 1959 to 2022. Note that the number of missions for 2022 is predicted and we assumed successful in all cases.

**2. Methodology**

2.1 NASA's Budget

While both the United States and the former USSR have invested significantly on space related activities since the late 1950s, a more complete historical evaluation of how budget affects extent and timing of missions can be gleaned from analyzing the readily accessible and reliable data from NASA. The U.S. has always attached importance to the industries which support space exploration and regards space development policy at the national level, thus the president directs the country's space activities while Congress is responsible for legislation, supervision and budget approval. NASA makes decisions on how best to meet directives within the constraints of imposed budgets and manages the execution of those decisions [9].

NASA's budget from 1959 to 2022, as depicted in Figure 2, notes an inflation-adjusted gradual rise. The trend can be easily plotted in a functional manner with the linear horizontal axis being the number of years since the start of the Space Age. A simple linear relationship is then best-fit to the plotted data, yielding:

$$B = 334.19T \qquad (1)$$

where B = NASA's budget in millions of dollars, T = time in number of years since 1957, and the correlation coefficient $R^2 = 0.9789$. Note the assumption that before the Space Age, NASA's budget was zero.

It should be noted that while Figure 1 utilizes inflation-adjusted budget data, Figure 2's budget data is left unadjusted for inflation. The adjustment in Figure 1 is intended to relate successful mission counts to budget, this on a year-by-year basis, revealing a time invariant connection between these two parameters. In order to compare different times, either specific years or decades, to each other along the 64-year scale of this plot, the time dependent change in purchasing power (i.e., inflation) must be normalized out. Hence, adjusting for inflation is necessary to expose the annual budget vs. mission count relationship. Figure 2, in contrast to Figure 1, illustrates the general trend of NASA's budget over that same time span. Here, the budget's variation with time is to be considered in isolation from all other factors such as inflation - in effect, simply having to pay more dollars for the same item or service at a later time - and technological capability, which is discussed and quantified on its own terms in Section 2.2.

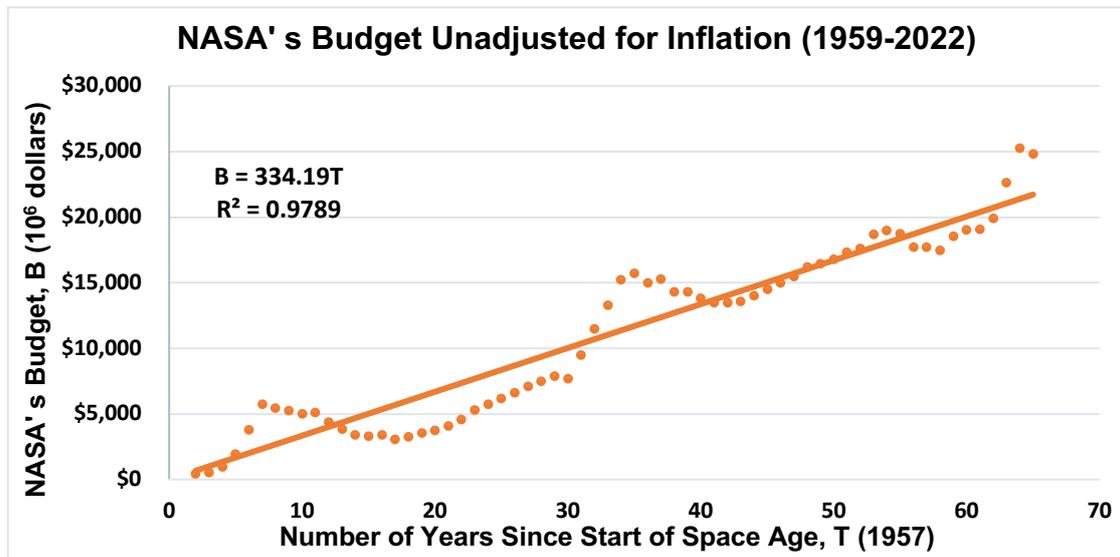

**Figure 2:** This graph shows NASA's budget from 1959 to 2022, time expressed as number of years since start of the Space Age (1957), used in empirically derived equation (1). NASA's budget data comes from https://www.planetary.org.

There are three small peaks in the actual NASA budget curve compared to the fitted line result. The first peak occurred when the United States and the Soviet Union were in the most intense portion of the Space Race. In 1966, when America's space activities were hitting their early stride as transition from the Gemini Program to the more ambitious Apollo Program began taking hold, the space budget accounted for about 1% of gross national product [10]. During this period the United States achieved fruitful results in deep space exploration. The peak in NASA's budget in the early 1990s was due to a decision to partner with private sector aviation/aerospace as soon as the space shuttle, extensively used for transporting astronauts and materials to Earth orbit, was gradually retired, in part owing to its space-bound cargo carrying capacity becoming insufficient. This new direction was made official in 1991 with President George H.W. Bush's signing of a new commercial space policy. The third peak, occurring in the final years of the 2010s, is the result of the September 2018 U.S. government release of the "National Space Exploration Activities Report" [14] which lays out a crewed exploration route from low Earth orbit to the Moon and then to Mars based on the Trump Administration's Space Policy Order 1 [11]. Thus, a careful analysis of NASA's budget trend reveals presidential policies, while having some effect, have not been significant enough to shift the budget in a sustained manner from otherwise steady growth described by historical linear fitting.

2.2 Development of Technological Capability

In considering the economic implications of human deep space exploration, there is an implicit assumption that technological development will keep to a sufficiently brisk pace so as to continue facilitating future research and development. Further, as the level of technology and productivity improve, this should be expected to drive down the cost of sophisticated and labor-intensive construction processes, attenuating the investment required for future crewed deep space missions.

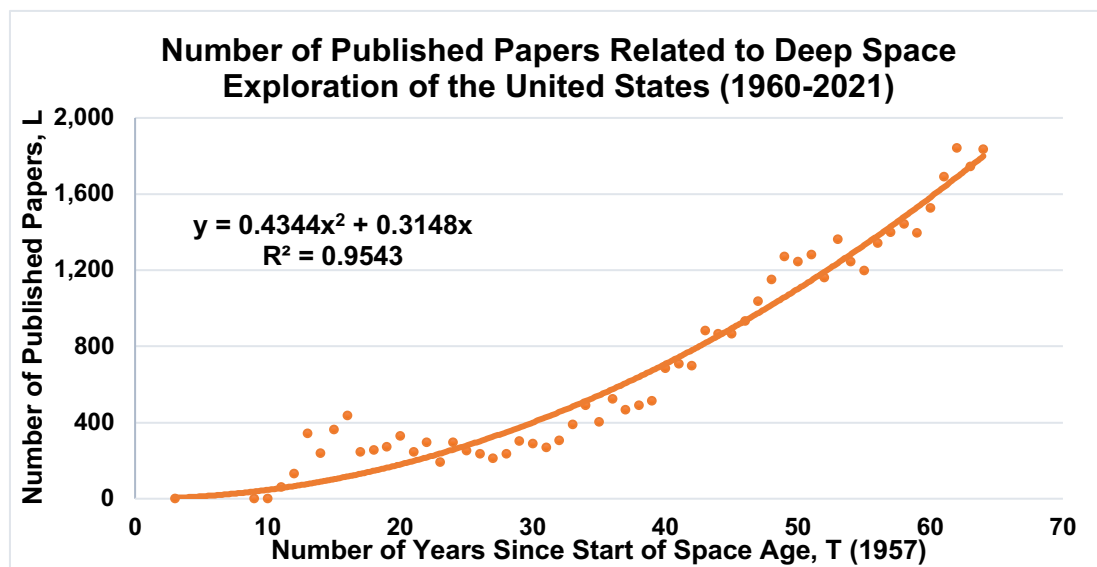

**Figure 3:** Shows a trend in deep space exploration technology level as expressed in number of related papers from start of the Space Age (1957), used in the empirically derived equation (2). Data sourced from https://www.webofscience.com/wos/alldb/summary/2fc00605-8724-4021-9056-ce253f047988-2a0f46c5/relevance/1 (accessed 05/13/2022)

Deep space exploration involves a variety of technologies such as computing power, which is leveraged in the design, manufacture, and operation of hardware, e.g., launch vehicles and guidance systems, as well as for life support where human crews are present. It is difficult to quantitatively describe the development of technology as a whole by the specifics of any one area of technology. However, as a first approximation the number of published papers on deep space exploration of the United States each year can be used as a proxy to gauge the overall technology level of cutting-edge developments in this complex realm. Directionally speaking, the greater the number of papers, the higher the (generalized) technology level rises and thus a scalable rate of development can be derived from this trend. Mapping the number of papers published per year (L) starting from the mid-1960s yields a dataset that can then be fitted by a quadratic function (Figure 3). As in Figure 2, the horizontal axis is taken as the number of years since the beginning of the Space Age. Assuming L = 0 for T = 0, the best fit result is:

$$L = 0.4344T^2 + 0.3148T \qquad (2)$$

2.3 Establishing the Model

To create our model, we define a new parameter R – the effective radius of human activity beyond Earth - which is taken as the minimal distance from Earth's orbit to a given objective's orbit and conditioned on a successful first human landing. For instance, there was a successful manned landing on the Moon in 1969, so the effective radius of human activity was 0.0026 AU in 1969. As crewed exploration advances, the effective radius of human activity will enlarge, hence R can be mapped to time T. These two parameters satisfy the following relationship:

$$\frac{dR}{dT} = ABL \qquad (3)$$

where A is a proportionality coefficient, L represents the technology level, and B represents NASA's budget. As NASA's annual budget, which generally grows even after adjusting for inflation, continues to fund deep space exploration, the effective radius of human activity continues to expand. The differential form of equation (3) reflects the overall time dependency of this characteristic as well. The linearly multiplicative term L can be interpreted as technological development effectively acting as a scalar on subsequent research, development and, ultimately, resulting in investment reduction if R were to remain fixed.

After substituting in the time dependent expressions for B and L, the parameter R can be obtained by integration of equation (3):

$$R = A(36.293T^4 + 35.068T^3) + C \qquad (4)$$

Here, the constant C is produced by the indefinite integration. There are two constants in this expression for R which can be solved for by plugging in actual and assumed data. For the first data point, we recall the first successful manned mission to the Moon in 1969, so for T = 12 years, R = 0.0026 AU. For the second data point an assumption is required as, even more than half a century later, the Moon remains the only celestial body visited by humans. While we have yet to step foot on Mars, the first crewed landing in the 2030s is being realistically envisioned. Acting administrator Steve Jurczyk of NASA's current Artemis Program has reiterated his aspiration of "the mid-to-end of the 2030s" for American boots on the Red Planet [12]. Although the applied energy optimizing Hohmann transfer orbit launch windows from Earth to Mars in the latter 2030s will actually occur in 2037 and 2039, for the purposes of this analysis we will pragmatically assume the average, thus the assumption for the first crewed mission to Mars successfully launching in 2038. The distance between Mars and Earth's orbit is 0.3763 AU, so for T = 81 years {2038 minus 1957}, R = 0.3763 AU. Using these two points, (12, 0.0026) and (81, 0.3763), the complete expression for R is rendered as:

$$R = 2.365 \times 10^{-10} \times (36.293T^4 + 35.068T^3) + 0.002 \qquad (5)$$

Again, T refers to the minimal number of years since 1957 for launch of human-crewed landing missions. Based on this model and equation (5), we can estimate the earliest launch timeframe for such missions to the outer reaches of the Solar System.

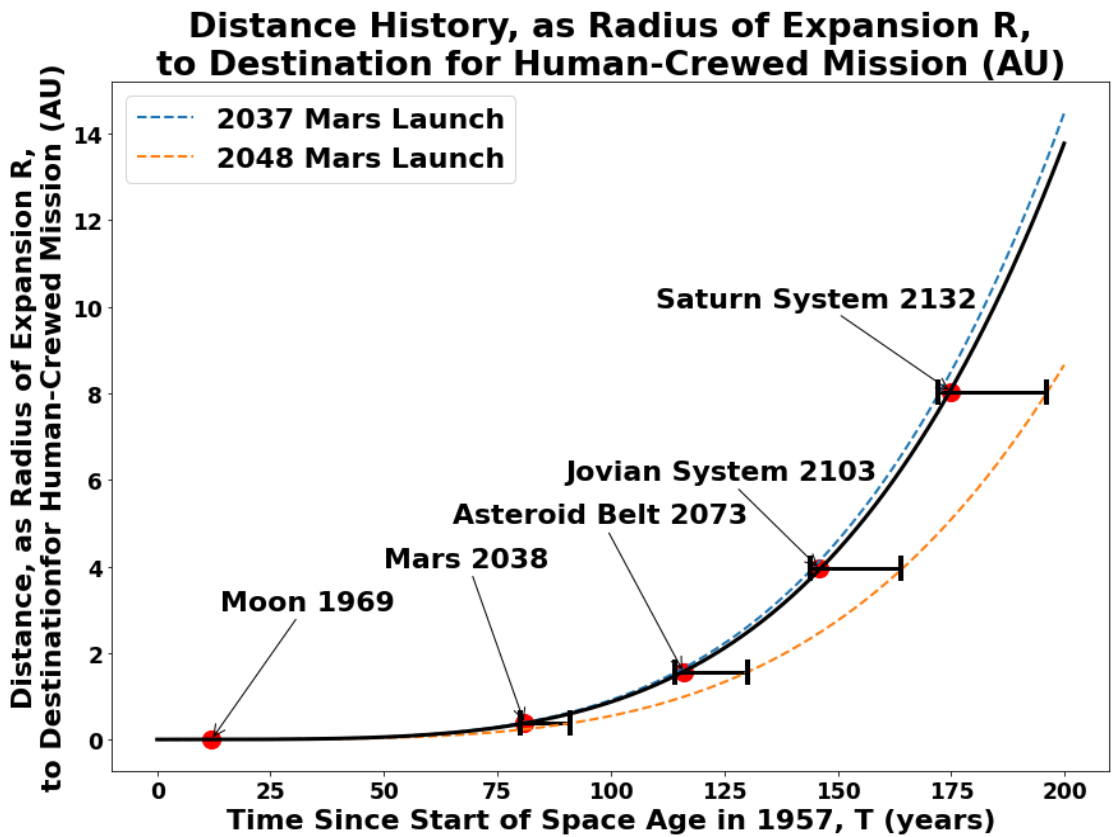

**Figure 4:** Shows the trend of effective radius of human activity as a function of time starting from the dawn of the Space Age (1957), projecting the earliest launch timing to various Solar System destinations and corresponding spread per assumed uncertainty in first successful crewed Mars mission launch date.

At this point it is necessary to consider and quantify the influence uncertainty exerts on our projections. This can be accomplished by assuming a reasonable variance in the launch year of the first crewed mission to Mars and its "knock-on" effect on missions to further destinations. While an aggressive schedule may envision such a launch by the late 2020s, this optimistic timing is deemed unlikely particularly given the challenging socioeconomic circumstances which have prevailed in recent years - e.g., the SARS-CoV-2 pandemic slowing research and development activities. A more prudent assumption would logically place the timing closer to the working range of 2037 to 2039, hence 2037 is conservatively chosen as the early launch date. Turning our attention to a delayed launch date for the first crewed Mars mission, we will assume an additional 10 years is needed to prepare. Given the overall duration of a crewed mission, which would be on the order of many months, substantial technological hurdles would have to be overcome in advance [14]. This condition opens the door to additional delays ahead of the crew's departure from Earth involving associated research, development and deployment. It is noted, however, that fully proven technological leaps in next generation human-rated vehicle developments have historically occurred with approximate decadal cadence – e.g., Mercury (1961) to Apollo (1969) to the Space Shuttle (1981). Accordingly, a late launch date for humans to Mars is placed in 2048. With the first successful crewed mission to Mars now bracketed as launching between 2037 and 2048, again with 2038 as effectively nominal, our parameters are complete and ready to be applied to equation (5). As illustrated in Figure 4, it may be possible to launch deep space explorers to the Asteroid Belt and Jovian System as soon as $2073^{+14}_{-2}$ and $2103^{+18}_{-2}$, respectively. Launch of a crewed mission to Saturn's moons may be realized in slightly over a century from now, circa $2132^{+21}_{-3}$.

The choice of destinations in the Solar System also takes into account the possibility of humanity ultimately colonizing these destinations in the more distant future, with long duration bases as those colonies' logical forerunners. Given the Moon is the closest celestial body to the Earth, and with current, proven spacecraft technology, astronauts have repeatedly traversed that distance in just four days; the next major step will be a lunar base. NASA's Lunar Exploration Program, which tentatively targets the south polar region of the Moon later in the 2020s, will land the first woman and first person of color. The establishment of a base on the Moon is envisioned to play an important role in preparation for subsequent crewed missions to Mars [13]. Mars is considered the next step per proximity, a choice supported by analysis suggesting it to be the most economically viable location in the Solar System for colonization including specific discussion about building a 1000-person Martian colony [6,7]. Given the findings of past and current robotic probes, and what future probes may uncover, the probability of humanity eventually coming to colonize selected locations in the Asteroid Belt, as well as some moons in the Jupiter and Saturn systems, is also conceivable [8].

## 3. Discussion of Future Work

The accuracy of predicting the earliest possible launch dates of human-crewed deep space missions could be further improved. There are two avenues to gain accuracy based on the methodology for deriving the projected results of this analysis.

Firstly, the accuracy of the prediction can be enhanced by a more detailed analysis of the U.S.'s investment in deep space exploration. We use NASA's all-inclusive budget trend to build the foundational model in this paper, but the budget is actually comprised of multiple components. The accuracy can thus be improved by subdividing that budget into higher resolution apportionments. Further layering the funding aspect are international cooperation and commercialization, for which the development trends are growing in the realm of deep space exploration, including direct human participation.

However, cross-border cooperation is greatly affected by the international environment and commercial companies are profit-oriented and thus subject to the fiduciary interests of their shareholders. Accordingly, these factors' contributions to promoting human deep space exploration pose challenges to weigh carefully alongside the analysis of well-documented government investment.

Secondly, and more broadly, there are many factors which affect the development of human deep space exploration, with the economic and associated research factors considered herein being just one. Related policies, number of employees and technical issues will all affect the earliest possible launch dates of human-crewed missions to the Solar System and beyond. Although the initial model built by previous work [8] projected a timeframe by considering computing power as the only factor affecting human deep space exploration, noted as well were other elements needing to be taken into account for a more complete picture to be drawn [8].

In conclusion, we can expect continued additions to the model of more far-ranging and precisely expressed influencing factors that will further improve the accuracy of the model's predicted results via the differential analysis method. Although preliminary, the results thus far suggest the worlds of our Solar System, throughout human history merely specs of light in the night sky, will soon be within our grasp.

## 4. Conclusion

A simple model based on trends built from NASA's budget and the development of human-crewed deep space exploration through the first six plus decades of the Space Age projects the earliest possible launch dates of crewed missions to distant destinations in the Solar System. Considering continued technological developments, in part driven by (assumed) ever-increasing computational power, will significantly and positively impact the future deep space exploration, a proxy for associated technical level influence is introduced as well. Due to the wide range of technologies involved in deep space exploration, and particularly the complexities of providing human life support across interplanetary distances, the trend in technological development is quantitatively represented by the trend in the number of related published papers.

While many would consider deep space exploration to be a noble endeavor in its own right, deserving of large investment by governments and private industry, such costs must also yield scientific and other value in return. Although varying economic conditions have and will continue to affect the pace of development of deep space exploration, our simplified model suggests the first human-crewed missions to land on selected Asteroid Belt objects could occur within the next half century and launches of human-crewed missions to selected moons of Jupiter and Saturn perhaps in the early 2100s. Although the Solar System is dauntingly large, human intelligence brought us from the first crude heavier-than-air flight in 1903 to the surface of the Moon only 66 years later – less than a current average human lifespan in developed nations. Our model, informed by data from prior decades, suggests human landings on worlds beyond the Moon and Mars may well be witnessed by many alive today. Of course, this result was obtained by accounting for economic factors based on their historical trend and the assumption of their continuation over the next 100+ years - which undeniably imposes uncertainties which themselves are difficult to quantify. As a counter, the resources, financial and otherwise, accorded deep space exploration are a strong function of national policy and, increasingly, private interest. We expect the resolve to keep pushing back the frontiers of space will continue to find the support needed to be successful.

**Acknowledgement:** This work was supported by the Jet Propulsion Laboratory, California Institute of Technology, under contract with NASA. We also thanks many supports from the department of astronomy of Beijing Normal University of China and department of computer sciences of Radboud University of The Netherlands.

**Data availability:** All data and software used for this study will be submitted online as attachments after peer-review. For additional questions regarding the data sharing, please contact the corresponding author at Jonathan.H.Jiang@jpl.nasa.gov.

## References

[1] Chen Daqing. History of human exploration of the moon (in Chinese) [J]. Foreign Space Dynamics, 1989 (7):2.
[2] Zheng Yongchun. A Brief History of Mars Exploration (in Chinese) [J]. Science.
[3] Wu Weiren, YU Dengyun. Journal of Deep Space Exploration (in Chinese), 2014, 1(1):13.
[4] LI Yifan. Research on key technologies of manned deep space exploration [J]. Missiles and Space Vehicles, 2018 (1):8.
[5] Zhang Yingyi, Zhang Wei. Status and development trend analysis of manned deep space exploration abroad (in Chinese) [J]. Space China, 2019 (11):6.
[6] Zubrin R. (2007) Economic Viability of Mars Colonization, Lockheed Martin Astronautics. Downloaded from: (https://web.archive.org/web/20070928081643/http://www.4frontierscorp.com/dev/assets/Economic%20Viability%20of%20Mars%20Conolozation.pdf), Accessed May 23, 2021.
[7] Vincenzo Donofrio, Meghan Kirk. Plan for Building a 1000 Person Martian Colony. arXiv:2112.06145
[8] Jiang J.H., Rosen, P.E., Fahy K.A, Avoiding the "Great Filter": A Projected Timeframe for Human Expansion Off-World, Galaxies, 9(3), 53, 2021.
[9] Luo Qinglang. Research on China's Manned Space Development and Budget Policy (in Chinese) [D]. Institute of Fiscal Science, Ministry of Finance, 2013.
[10] Shen Qing. Review and Prospect of American Space Funds (in Chinese) [J]. Foreign Missiles and Astronautics, 1984 (03):35-37+18.
[11] Zhang Luyun, Qu Jing. The development of Manned spaceflight in the United States from NASA budget for fiscal year 2020 (in Chinese) [J]. Aeronautical Missile, 2020 (2):6.
[12] Ahmed I., Aubourg L. (2021) America has Sent Five Rovers to Mars – When Will Humans Follow? Phys.org. Downloaded from: (https://phys.org/news/2021-02-america-roversmarswhen-humans.html). Accessed May 17, 2021.
[13] NASA's Lunar Exploration Program Overview. Downloaded from https://www.nasa.gov/specials/artemis/
[14] National Space Exploration Campaign Report. Downloaded from: https://www.nasa.gov/sites/default/files/atoms/files/nationalspaceexplorationcampaign.pdf
[15] National Space Exploration Campaign Report. Downloaded from: https://nssdc.gsfc.nasa.gov/nmc/spacecraft/display.action?id=1969-043A.